\documentclass[conference]{IEEEtran}
\usepackage{float}
\usepackage{subfigure}
\usepackage{caption}
\usepackage{cite}
\usepackage{graphicx}
\usepackage{amsmath}
\usepackage{mathtools}
\usepackage[table,xcdraw]{xcolor}
\usepackage{booktabs}
\newcommand{\squeezeup}{\vspace{-2.5mm}}
\newcommand{\squeezeupt}{\vspace{-1.5mm}}
\ifCLASSINFOpdf
\else
\fi
\begin{document}
%
\title{FastWave: Accelerating Autoregressive Convolutional Neural Networks on FPGA}



%
\author{\IEEEauthorblockN{Shehzeen Hussain\IEEEauthorrefmark{1},
Mojan Javaheripi\IEEEauthorrefmark{1},
Paarth Neekhara\IEEEauthorrefmark{2}, 
Ryan Kastner\IEEEauthorrefmark{2} and
Farinaz Koushanfar\IEEEauthorrefmark{1}}
\IEEEauthorblockA{\IEEEauthorrefmark{1}UC San Diego Department of Electrical and Computer Engineering}
\IEEEauthorblockA{\IEEEauthorrefmark{2}UC San Diego Department of Computer Science\\
Email:  ssh028@eng.ucsd.edu, mojavahe@eng.ucsd.edu}
}


\maketitle

\thispagestyle{plain}
\pagestyle{plain}
\def\edit{\textcolor{blue}}
\def\comment{\textcolor{magenta}}

\begin{abstract}
Autoregressive convolutional neural networks (CNNs) have been widely exploited for sequence generation tasks such as audio synthesis, language modeling and neural machine translation. 
WaveNet is a deep autoregressive CNN composed of several stacked layers of dilated convolution that is used for sequence generation. While WaveNet produces state-of-the art audio generation results, the naive inference implementation is quite slow; it takes a few minutes to generate just one second of audio on a high-end GPU. In this work, we develop the first accelerator platform~\textit{FastWave} for autoregressive convolutional neural networks, and address the associated design challenges. We design the Fast-Wavenet inference model in Vivado HLS and perform a wide range of optimizations including fixed-point implementation, array partitioning and pipelining. Our model uses a fully parameterized parallel architecture for fast matrix-vector multiplication that enables per-layer customized latency fine-tuning for further throughput improvement. Our experiments comparatively assess the trade-off between throughput and resource utilization for various optimizations. Our best WaveNet design on the Xilinx XCVU13P FPGA that uses only on-chip memory, achieves $\mathbf{66\times}$ faster generation speed compared to CPU implementation and $\mathbf{11\times}$ faster generation speed than GPU implementation. 
\end{abstract}


%
\IEEEpeerreviewmaketitle


\section{Introduction}

 


Autoregressive convolutional models achieve state-of-the-art results in audio ~\cite{wavenet,tts,deepvoice2,Qian2017SpeechEU} and language domains ~\cite{bytenet,vae} with respect to both estimating the data distribution and generating high-quality samples. 
Wavenet~\cite{wavenet} is an example of autoregressive convolutional network, used for modelling audio for applications such as text-to-speech (TTS) synthesis and music generation.
WaveNet has been rated by human listeners to provide substantially more natural sounding audio when compared to the best existing parametric and concatenative systems in TTS applications for both English and Mandarin\cite{wavenet}. Popular cloud based TTS synthesis systems such as Google Now and Google Assistant, that produce natural sounding speech, are built on WaveNet architecture~\cite{Qian2017SpeechEU,deepmind}. Alongside achieving state-of-the art results in the audio domain, convolutional models are prominent for natural language modeling tasks like text generation and machine translation~\cite{bytenet}.

Generally, both autoregressive convolutional neural networks (CNNs) and Recurrent Neural Networks (RNNs) ~\cite{rnn_paper} are widely popular for sequence modelling tasks.  The main advantage of CNN based models is that they can achieve higher parallelism during training and can capture longer time-dependencies as compared to RNN based models~\cite{empirical_comparison,efficienttts_attention}. However, this comes at a cost of slower inference, since the inference still remains sequential and CNN based models are usually very deep. However, to overcome this problem, Fast-Wavenet~\cite{fastwavenet} exploits the temporal dependencies by caching redundant computations using fixed-length convolutional queues and thus makes the generation time linear in length of the sequence. Such efforts have made it feasible to use autoregressive CNNs for practical sequence generation applications, as an alternative to RNN-based models. 

While the Fast-Wavenet algorithm provides a speed-up in audio generation over na\"ive Wavenet implementation, the generation time is still high, even on a high-end GPU. It takes 120 seconds to generate 1 second of audio using Fast-Wavenet implementation on a NVIDIA TITAN Xp GPU. Prior works have shown FPGAs to be successful in accelerating the inference of pre-trained neural networks by providing custom data paths to achieve high parallelism. A vast amount of such research focuses on accelerating neural networks in the image domain~\cite{samragh2017customizing,sharma2016high}, speech recognition~\cite{fpga_lowpower,speechrecfpga} and language modelling~\cite{rnnlanguage}. To the best of our knowledge, similar efforts have not been made for accelerating neural networks for speech/audio synthesis. 

We aim to accelerate audio synthesis using the autoregressive CNN model - WaveNet on FPGA. The primary challenges in deploying auto-regressive CNN inference on FPGA are designing modules for dilated convolutional layers, buffers for storing redundant computations using convolutional queues, and dealing with the large depth of these networks which is necessary to maintain high audio quality. 
In this work, we address these challenges of deploying large autoregressive convolutional models on FPGAs and perform a wide range of application and architectural optimizations. Furthermore, we comprehensively analyze and compare the performance of Fast-Wavenet implementation on FPGA with the CPU and GPU counterparts.


\noindent \textbf{Summary of Contributions:}
\begin{itemize}
    \item Creation of the first accelerator platform for autoregressive convolutional neural networks. We deploy the fast inference model Fast-Wavenet\cite{fastwavenet} on Xilinx XCVU13P FPGA which achieves \textbf{11} times faster generation speed than a high-end GPU and \textbf{66} times faster generation speed than a high-end CPU.

    \item Development of reconfigurable basic blocks pertinent to autoregressive convolutional networks i.e., dilated causal convolutional layers,  convolutional queues, and fully connected layer. Our operations are powered by a fully-customizable matrix-multiplication engine that uses two levels of parallelism controlled by tunable parameters.
    
    \item Creation of an end-to-end framework that accesses only on-chip memory thereby ensuring high throughput and avoiding any latency caused by communication with off-chip memory. 
    
    \item Exploration of the design space using different architectural and application optimizations, as well as comparing the performance and resource usage. We present extensive evaluation of throughput and power efficiency for our fully optimized and baseline designs.
    
    
\end{itemize}



\section{Background}\label{sec:background}
In this section, we provide a background on autoregressive convolutional neural networks. We choose WaveNet as an ideal representation of such models, and describe its overall generative architecture. We first elaborate on the 1D convolution operation as it is the core computation performed in the WaveNet model. Next, we explain WaveNet and its more efficient inference algorithm called  Fast-Wavenet.

\subsection{\textbf{1D Convolution}}\label{sec:1d_conv}
The 1D convolution operation is performed by sliding a one dimensional kernel over a 1D input signal. Each output value at position $i$ is produced by calculating the dot product of the kernel and the overlapping values of the input signal, starting from position $i$. More formally, for an input vector $a$ of length $n$ and a kernel $k$ of length $m$, the 1D convolution is calculated as follows:
\begin{equation}
    (a*k)_i = \sigma_{j=1}^m k_j\times a_{i-j+\frac{m}{2}}
\end{equation}
where i is an arbitrary index in the output vector, which has a total length of $n-m+1$. The subscripts denote the indices of the kernel/input vectors. 


\subsection{\textbf{WaveNet and Autoregressive CNNs}}\label{sec:wavenet}
Autoregressive Neural Networks are popularly used for sequence generation tasks which rely on ancestral sampling i.e. the predictive distribution for each sample in the sequence is conditioned on all previous ones. While RNNs are popular autoregressive models, they do not exhibit high receptive field making them unsuitable for modeling sequences with long-term dependencies like audio \cite{empirical_comparison}. In contrast, autoregressive CNN based models use a stack of dilated convolutional layers to achieve higher receptive field necessary for modeling sequences with long-term dependencies.   

Wavenet~\cite{wavenet} is an autoregressive convolutional neural network that produces raw audio waveforms by directly modeling the underlying probability distribution of audio samples. This has led to state-of-the-art performance in text-to-speech synthesis~\cite{tts,deepmind,Tamamori2017SpeakerDependentWV,liu2018wavenet}, speech recognition~\cite{speech_recognition_wavenet}, and other audio generation settings~\cite{wavenet,deepvoice2,Qian2017SpeechEU}. The Wavenet architecture aims to model the conditional probability among subsequent audio samples. The joint probability distribution of waveform sample points $x =  {x_0, x_1, ..., x_T}$ can be written as:
$P(x | \lambda) = \prod_{t=1}^T P(x_t | x_{t-1}, .., x_0, \lambda)$
where $\lambda$ denotes the learnable parameters of Wavenet model.
During inference, next-sample audio ($x_t$) generation is performed by sampling from the conditional probability distribution given all of the previous samples, $P(x_t|x_{t-1}, ..., x_1, x_0, \lambda)$. 

\begin{figure}[htp]
    \centering
    \subfigure[]{\includegraphics[width=0.7\columnwidth]{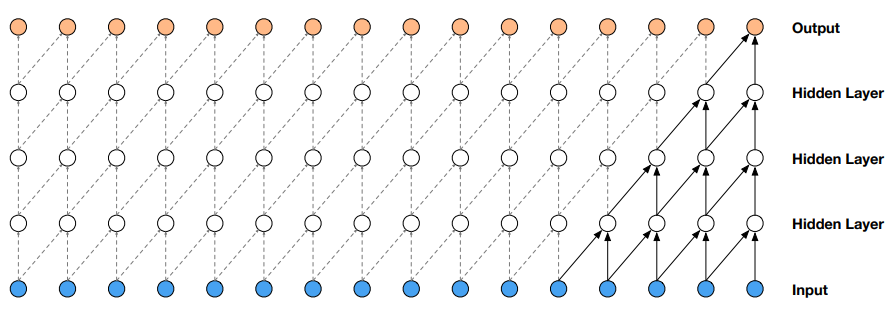}\label{fig:causal_conv}}
    \subfigure[]{\includegraphics[width=0.7\columnwidth]{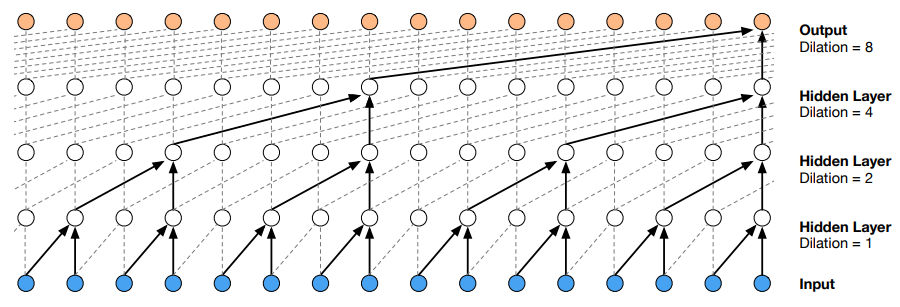}\label{fig:dilated_conv}}
    \caption{\textbf{a.} (Left) Stacked causal convolution layers without any dilations. \textbf{b.} (Right) Stacked causal 1-d convolution layers with increasing dilation. (Figures from WaveNet paper~\cite{wavenet}).}
\end{figure}

One possible method for modeling the probability density is via a stack of causal convolutional layers as depicted in Figure~\ref{fig:causal_conv}. The input passes through this stack of convolutional layers and gated activation functions and finally through a softmax layer to get the posterior probability  $P(x_t|x_{t-1}, ..., x_1, x_0)$.
The downside of this approach is that in order to model long temporal dependencies from samples far in the past, the causal convolutional network requires either many layers or large filters to increase the receptive field. In general, the receptive field is calculated as $\#~of~layers + filter_{size}+1$ which gives a receptive field of 5 in the architecture shown in Figure~\ref{fig:causal_conv}. To address this problem, WaveNet leverages dilated convolutions~\cite{yu2015multi,chen2018deeplab} which deliver higher receptive fields without significant increase in computational cost of the model. Dilated convolution is equivalent to performing convolutions with dilated filters where the size of the filter is expanded by filling the empty positions with zeros. In practice, this is achieved by performing a convolution where the filter skips input values with a certain step. 

Fig~\ref{fig:dilated_conv} illustrates a network with dilated causal convolutions for dilation values of 1, 2, 4, and 8. Here, the input nodes are shown with color blue and the output is shown with orange. Each edge in the graph correspond to a 1-dimensional convolution (See section~\ref{sec:1d_conv}), more generally a matrix multiplication. Due to the binary tree structure of the network, the time complexity of computing output at each time-step is $\mathcal{O}(2^L)$ where $L$ is the number of layers in the network, which gets highly undesirable as $L$ increases.  Similarly, the total memory required to store the inputs, output, and the intermediate layer features is $\mathcal{O}(2^L)$.

\subsection{\textbf{Fast-Wavenet}}\label{fastwavenetsec}
The na\"ive implementation in Figure~\ref{fig:dilated_conv} has many redundant computations when generating a new sample, that is, it recomputes activations that have been already computed for generating previous samples. Fast-Wavenet \cite{fastwavenet} proposed an efficient algorithm that caches these recurrent activations in queues instead of recomputing them from scratch while generating a new sample. Fast-Wavenet uses a per-layer first-in-first-out queue to cache the states to be used in future timestamps. 

\begin{figure}[H]
\centering
\includegraphics[height=1.3in]{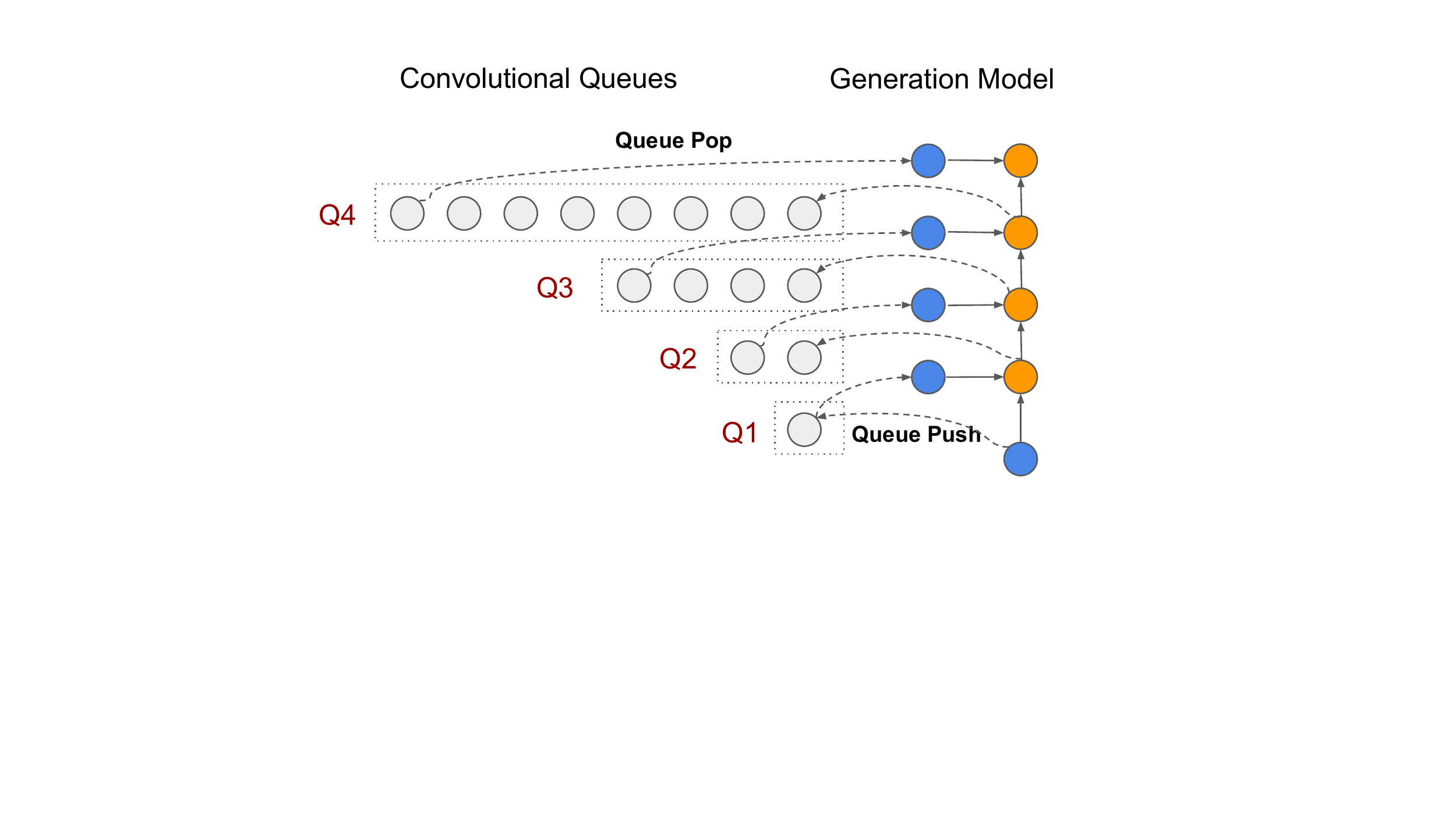}
\caption{Basic queue operations (Push and Pop) performed in Fast-Wavenet to achieve linear time in audio generation.}
\label{fig:pop_phase}
\end{figure}

The queue size at each layer is determined by its corresponding dilation value. Figure~\ref{fig:pop_phase} demonstrates an example $4$-layer network and their corresponding queues. For the first layer, the dilation value is $1$ and therefore the corresponding queue~($Q1$) only keeps one value. Similarly, the output layer has a dilation value of 8, which means that its queue~($Q4$) will store 8 recurrent values. By removal of redundant computations due to the queue storing mechanism, the computational complexity of Fast-Wavenet is $\mathcal{O}(L)$ where $L$ is the number of layers. The overall memory requirement for queues as well as the  intermediate values remains the same as the na\"ve implementation, i.e., $\mathcal{O}(2^L)$. 

The basic queue operations performed in the Fast-Wavenet are as follows (refer to Figure ~\ref{fig:pop_phase} ):
\begin{enumerate}
    \item \textbf{Pop phase}: The oldest recurrent states are popped off the queues in each layer and fed as input to the generation model. These popped off states and the current input are operated with the convolutional kernel to compute the current output and the new recurrent states.
    \item \textbf{Push Phase}: Newly calculated recurrent states~(orange dots) are pushed to the back of their respective layer queues to be used in future time stamps.
\end{enumerate}

\noindent Maintaining the convolutional queues in the above manner allows us to handle the sparse convolutional operation and  avoid redundant computations and makes the generation algorithm linear in terms of length of the sequence.

\section{Methodology}

Our primary objective is to accelerate the inference of autoregressive CNNs for sequence generation on FPGAs. As an ideal candidate for autoregressive models, we choose the WaveNet model for raw audio generation from random seed inputs. The computation and storage complexity of such state-of-the-art autoregressive CNNs is very high, particularly our FastWave architecture comprises of 28 convolutional layers with 128 channels each in order to maintain high audio quality. When designing an accelerator for such models, it is important to be aware of the system restrictions, particularly those of memory access bandwidth ~\cite{zhang2015optimizing,samragh2017customizing}. Accessing off-chip memory is expensive and can limit the throughput of our network, making it important to compress DNNs into an optimal model for efficient inference. 

\textbf{Design Flow:} We start with an open source TensorFlow implementation of the Fast-Wavenet algorithm. We save the weights of the convolutional and fully connected layers of our trained model which are used in the inference stage for generating audio. We implement the Fast-Wavenet inference in NumPy without using any high level deep learning libraries. This implementation serves as a bridge between the high level TensorFlow and the low level Vivado HLS implementation in C++. On the FPGA platform, we then accelerate the audio generation process from random seeds, and perform optimized operations using queue buffers and matrix-vector multiplications to generate raw audio. 



\subsection{\textbf{Model Architecture and Training on GPU}}

\begin{table}[h]
\centering
\resizebox{0.85\columnwidth}{!}{%

\begin{tabular}{@{}ccccccc@{}}
\toprule
\textbf{\begin{tabular}[c]{@{}c@{}}Block \\ No.\end{tabular}} & \textbf{\begin{tabular}[c]{@{}c@{}}Layer \\ No.\end{tabular}} & \textbf{\begin{tabular}[c]{@{}c@{}}Filter \\ Width\end{tabular}} & \textbf{\begin{tabular}[c]{@{}c@{}}Queue \\ Length\end{tabular}} & \textbf{\begin{tabular}[c]{@{}c@{}}Input \\ Channels\end{tabular}} & \textbf{\begin{tabular}[c]{@{}c@{}}Output \\ Channels\end{tabular}} & \textbf{\begin{tabular}[c]{@{}c@{}}Queue \\ Size\end{tabular}} \\ \midrule
1 & 1 & 2 & 1 & 1 & 128 & 1 \\
1 & 2 & 2 & 2 & 128 & 128 & 256 \\
1 & 3 & 2 & 4 & 128 & 128 & 512 \\
1 & 4 & 2 & 8 & 128 & 128 & 1024 \\
1 & 5 & 2 & 16 & 128 & 128 & 2048 \\
1 & 6 & 2 & 32 & 128 & 128 & 4096 \\
1 & 7 & 2 & 64 & 128 & 128 & 8192 \\
1 & 8 & 2 & 128 & 128 & 128 & 16384 \\
1 & 9 & 2 & 256 & 128 & 128 & 32768 \\
1 & 10 & 2 & 512 & 128 & 128 & 65536 \\
1 & 11 & 2 & 1024 & 128 & 128 & 131072 \\
1 & 12 & 2 & 2048 & 128 & 128 & 262144 \\
1 & 13 & 2 & 4096 & 128 & 128 & {\color[HTML]{CB0000} 524288} \\
1 & 14 & 2 & 8192 & 128 & 128 & {\color[HTML]{CB0000} 1048576} \\
2 & 1 & 2 & 1 & 128 & 128 & 128 \\
2 & 2 & 2 & 2 & 128 & 128 & 256 \\
2 & 3 & 2 & 4 & 128 & 128 & 512 \\
2 & 4 & 2 & 8 & 128 & 128 & 1024 \\
2 & 5 & 2 & 16 & 128 & 128 & 2048 \\
2 & 6 & 2 & 32 & 128 & 128 & 4096 \\
2 & 7 & 2 & 64 & 128 & 128 & 8192 \\
2 & 8 & 2 & 128 & 128 & 128 & 16384 \\
2 & 9 & 2 & 256 & 128 & 128 & 32768 \\
2 & 10 & 2 & 512 & 128 & 128 & 65536 \\
2 & 11 & 2 & 1024 & 128 & 128 & 131072 \\
2 & 12 & 2 & 2048 & 128 & 128 & 262144 \\
2 & 13 & 2 & 4096 & 128 & 128 & {\color[HTML]{CB0000} 524288} \\
2 & 14 & 2 & 8192 & 128 & 128 & {\color[HTML]{CB0000} 1048576} \\ \bottomrule
\end{tabular}%
}

\caption{Details of Fast-Wavenet Architecture. The column \textit{Queue Size} denotes the number of floating point numbers stored in each queue and is equal to $QueueLength \times InputChannels$.
}
\label{fig:arch}
\end{table}

We use an open-source TensorFlow implementation of Fast-Wavenet \cite{fastwavenet} to pre-train our network in Python. 
The network architecture we use is a stack of two dilated convolutional blocks. Each block consists of 14 convolutional layers with kernel size (filter width) = 2 and dilation increasing in powers of 2. Therefore each of the kernels is a 3-dimensional array of shape $2 \times input channels \times output channels$. The number of output channels in each layer is 128 in the baseline implementation. After each convolutional layer there is a $\mathit{tanh}$ activation function which serves as the non-linearity in our model as used in the original WaveNet paper \cite{wavenet}. A $tanh$ activation normalizes values between -1 and 1 and also allows us to better utilize fixed point data-types in the Vivado implementation without compromising on accuracy.

After the two convolutional blocks, we have a single fully connected layer which maps the activation of size 100 from the last convolutional layer to an output vector of size 256 followed by a softmax normalization layer. The output after the softmax layer is the generated distribution. The target audio is quantized linearly between -1 and 1 into 256 values. The one-hot representation of each sample of size 256 serves as the target distribution at each time-step. The cross entropy loss between the generated and target distribution is back-propagated to train the convolutional kernels and weights of the fully connected layer. 
\textbf{Memory challenges:} The primary memory bottle-neck in implementing the Fast-Wavenet inference is not the parameters of the convolutional kernels, but convolutional queues which cache the intermediate outputs of the convolutional layers to be used for future predictions. The size of these queues increases exponentially with the depth of the block in the neural network. As highlighted in Table \ref{fig:arch}, the 14th convolutional queue in each of the blocks stores $1,048,576$ floating point numbers $(\approx 33Mb)$. One way of addressing this challenge is to reduce the number of channels in the 13th and 14th convolutional layers via pruning. However in our experiments, we found pruning to degrade the quality of generated audio. To address this memory challenge without pruning the network, we utilize both BRAMs and URAMs available on our FPGA board. We store all convolutional queues on the BRAMs by default and off-load the 14th convolutional queue of each block onto the URAMs on our board. In this way, we are able to utilize only on-chip memory and achieve higher bandwidth without compromising on audio quality.



\subsection{\textbf{Accelerator Design Overview}}
\begin{figure}[htp]
\centering
\includegraphics[width=0.95\columnwidth]{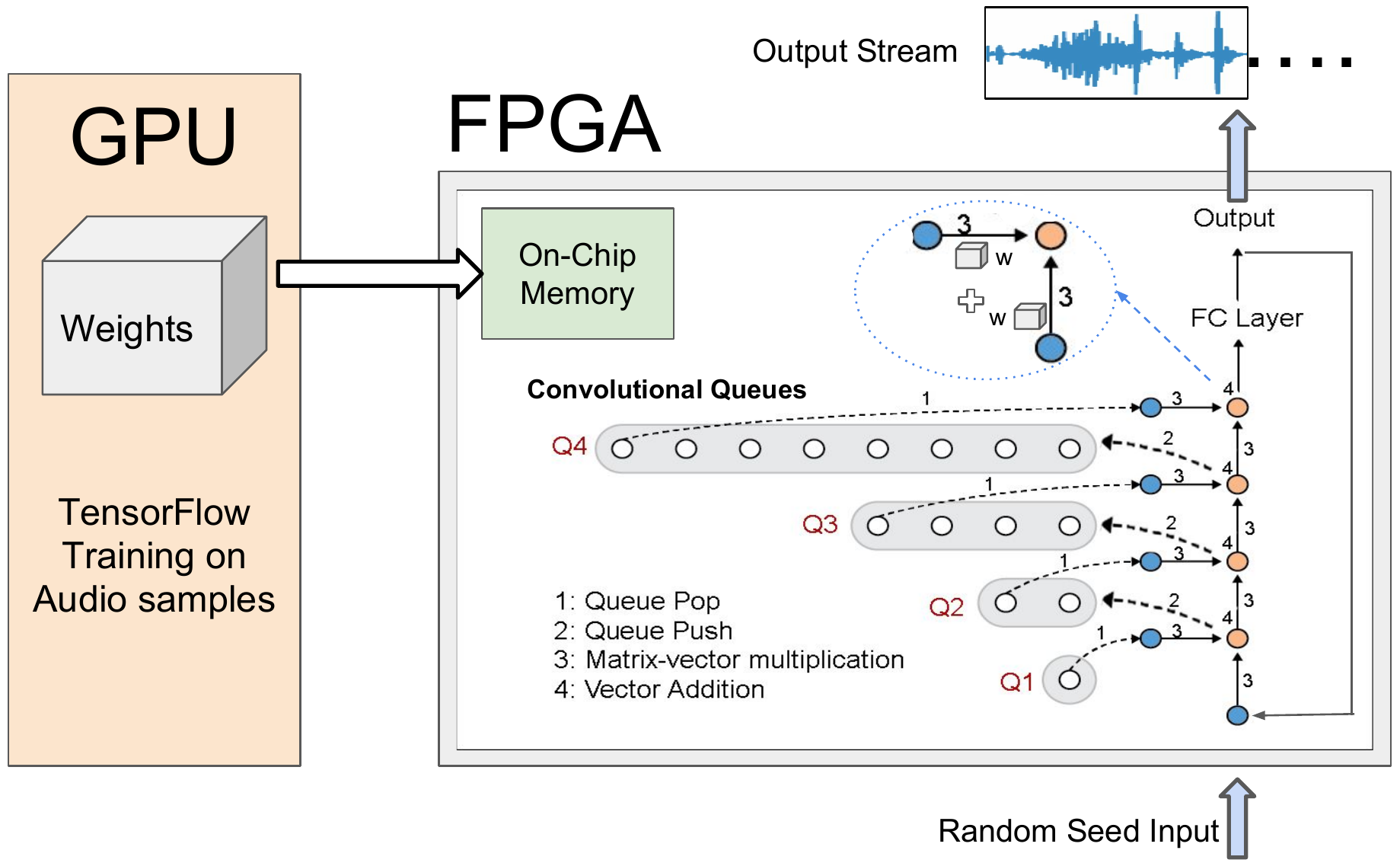}
\caption{Acceleration Methodology}\label{fig:design_overview}
\end{figure}
The primary objective of our system is to generate an output stream given a seed input. Figure \ref{fig:design_overview} shows the overview of our accelerator design. Given a seed input, our system generates an output stream in an autoregressive manner, one-sample at a time. The output sample produced at each time-step is fed back as input to generate the next output sample. During each cycle, as the input goes through all the convolutional layers, the corresponding convolutional queues are updated using push and pop operations as explained in section \ref{sec:background}. It is important to note that the entire model including the convolutional queues and the parameters does not use any off-chip memory and are stored in the BRAM and URAM available on the FPGA board. We describe the details of implementing the convolution operations, queue updates and output generation using the fully connected layers in the following section.



\section{Implementation Details}

Our design is composed of $5$ main elements: (i) The \textit{dilated convolution} layers, (ii) \textit{the queue control unit}, (iii) the \textit{fully-connected} layer,  (iv) the \textit{matrix multiplication engine}, and (v) the \textit{network description module}. We implement and accelerate the inference of WaveNet on the Xilinx XCVU13P FPGA.

\subsection{\textbf{Dilated Convolution Layer}}\label{sec:diconv}
As explained in the Section~\ref{fastwavenetsec}, Fast-WaveNet leverages queues to implement the dilated convolutional layers. A convolution of size $=2$ is used in the WaveNet architecture, and can be implemented as two matrix-vector multiplications followed by vector addition in the manner explained below. Notations used for our variables along with the shapes are listed below:\\\\
$IC_n$ : Number of Input channels of layer n.\\
$OC_n$ : Number of Output channels of layer n.\\
$O[n]_{( OC_n \times 1)}$ : Output of convolutional layer n.\\
$K[n]_{(2 \times OC_n \times IC_n)}$ : Convolutional kernel of layer n.\\
$Q[n]_{(queueLength \times IC_n)}$ : Convolutional queue of layer n.\\
$$O_1[n] = K[n][0]_{(OC_n \times IC_n)} \times Q[n][0]_{(IC_n \times 1)} $$
$$O_2[n] = K[n][1]_{(OC_n \times IC_n)} \times O[n-1]_{(IC_n \times 1)} $$
$$O[n]_{OC_n \times 1} = O_1[n]_{(OC_n \times 1)} + O_2[n]_{(OC_n \times 1)}$$

In other words, we matrix-multiply the first component of the convolutional kernel with the first element of the queue, and the 2nd component of the kernel with the previous layer's output and then add the two products to obtain the output of any layer. The details of the matrix-vector multiplication engine have been provided in Section {\ref{sec:matmul}}.

The output of the convolution layer is then passed to $tanh$ activation function. We use the CORDIC implementation available in Vivado HLS math library for applying $tanh$ allowing us to optimize our design and memory usage. The output of the dilated convolution module is a vector of length equal to the number of layer output channels.

\subsection{\textbf{Cyclic Queue Buffer Unit}}\label{sec:queue_control}
In order to reduce the number of operations, Fast-Wavenet aims to remove redundant convolution operations by caching previous calculations in a Queue, thereby reducing the complexity of synthesis to O(L) time. This means that after performing a convolutional operation, we push the compute into the end of the queue and pop the out the first element. These push and pop operations are shown in figure \ref{fig:design_overview}. As described above the queue in each layer $Q[n]$ is a 2-d array of shape $QueueLength \times Input Channels$. The $QueueLength$ depends on the $dilationFactor$ of the layer and is equal to $2^{dilationFactor}$. We aim to fit our queue computations in the on-chip memory BRAMs and URAMs. Our baseline queue implementation in Vivado HLS used shift operations to perform pop and push functionalities of a queue. The longest queue in our model is of size $8192 \times 24$. The shifting of a large number of elements in the queue resulted in very high latency. 

To make queue push and pop operations computationally efficient, we implemented our queues using fixed length circular arrays for each layer. This is a lot more efficient than shifting all the elements present in the queue. The push and pop operations are reduced to just overwriting one column of our circular array which is indexed using modulo $QueueLength$ index. 


\subsection{\textbf{Fully-connected Layer}}\label{sec:fc_layer}
The fully connected layer in WaveNet is a linear layer after all the convolutional layers. This layer is characterized by a weight matrix $W_{channels \times OutputSize}$ and a bias vector $b_{1 \times OutputSize}$. The fully connected layer performs the following operation on $ConvOut$: the output of the last convolution layer:
$$FinalOutput = ConvOut \times W  + b$$

In our design, the weight matrix $W$ has shape $100 \times 256$ and bias $b$ has shape $1 \times 256$.
We use arg-max sampling on the final vector of length $256$ to obtain the quantized output value between -1 and 1.

\subsection{\textbf{Matrix Multiplication Engine}}\label{sec:matmul}
\begin{figure}[htp]
    \centering
    \includegraphics[width=0.8\columnwidth]{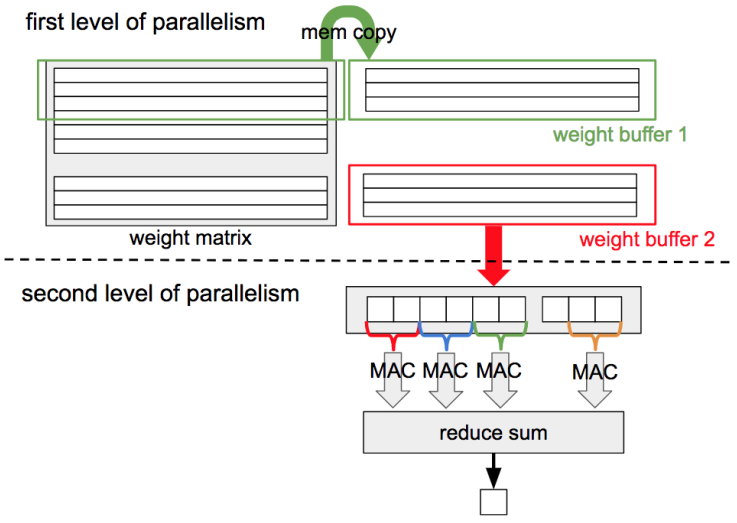}
    \caption{Schematic representation of the matrix multiplication engine and the corresponding parallelization factors.}\label{fig:matmul}
\end{figure}

The most computationally-intensive operation in DNN execution is matrix-vector multiplication. FPGAs are equipped with DSP units which offer a high computation capacity together with the reconfigurable logic. The basic function of a DSP unit is Multiplication Accumulation (MAC). Layers in a convolutional neural network take as input a vector
$X_{N\times1}$ and compute the output vector $Y_{M\times1}$ as formulated below:
\begin{equation}\label{eq:dot-product}
    Y = f(WX + b)
\end{equation}
where f(.) is a nonlinear function, $W_{M\times N}$ is the 2D matrix of the weights and $b_{M\times1}$ is a vector of bias values. As can be seen, each layer is computing a vector-matrix multiplication and a vector-vector addition. In order to optimize the design and make efficient use of the DSP blocks, we proposed a parallelized approach to convert layer computations into multiple MAC operations. Figure~\ref{fig:matmul} presents our method to parallelize the matrix-vector multiplication computations. 

\begin{figure}[htp]
    \centering
    \includegraphics[width=0.55\columnwidth]{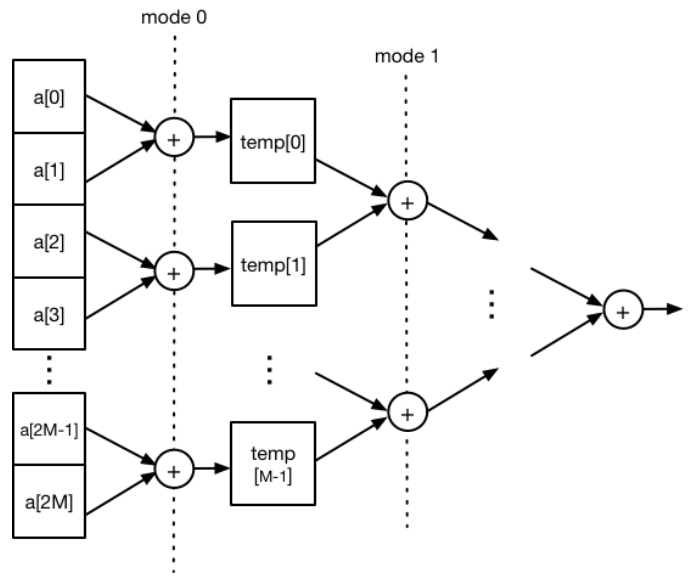}
    \caption{Realization of the tree-based vector reduction algorithm.}\label{fig:tree}
\end{figure}

We define two levels of parallelism for our engine which control the parallel computations with parameters \textit{num\_parallel\_in} and \textit{num\_parallel\_out}, denoting the level of parallelism in the input and output, respectively. For the first level of parallelism, multiple rows of the weight matrix are processed simultaneously by dividing it into chunks, each having \textit{num\_parallel\_out} rows. In each round, a chunk of the weights matrix is copied to one of the weight buffers while the other weight buffer is fed into the \textit{dot product} modules together with a copy of the input vector. The iterations end when all rows of the weight matrix have been processed. For the second level of parallelism,
each \textit{dot-product} function partitions its input vectors into \textit{num\_parallel\_in} chunks and concurrently executes MAC operations over the partitioned subsets. The accumulated results of the subsets are then added together within the \textit{reduce\_sum} function to compute the final output. The \textit{reduce\_sum} module performs a tree-base reduction algorithm as outlined in Figure~\ref{fig:tree}. The reduction function takes an array of size $2M$ as its input (array \textit{a}) and oscillates between 2 different modes. In mode $0$, the function reduces \textit{a} by using temp as a
temporary array. In mode $1$, temp is reduced using \textit{a}. The result is returned based on the final mode.

The aforementioned parameters \textit{num\_parallel\_in} and \textit{num\_parallel\_out} are individually defined for each of the layers to enable fine-tuning according to the per-layer requirements. Due to the limited number of available resources on the FPGA platform, it is not possible to define high parallelization factors for all layers. As such, we give priority to layers with higher computational complexity, i.e., higher number of input and output channels, by instantiating their corresponding matrix multiplication engines with larger parallelization parameters.

\begin{table*}[htp]
\resizebox{\textwidth}{!}{%
\begin{tabular}{|l|c|c|c|c|c|c|c|c|c|c|}
\hline
 & \multicolumn{5}{c|}{\textbf{Resource Utilization}} & \multicolumn{3}{c|}{\textbf{Performance}} & \multicolumn{2}{c|}{\textbf{Correctness}} \\ \hline
 & \textbf{\begin{tabular}[c]{@{}c@{}}BRAM \\ (Mb)\end{tabular}} & \textbf{\begin{tabular}[c]{@{}c@{}}URAM \\ (Mb)\end{tabular}} & \textbf{\begin{tabular}[c]{@{}c@{}}FF\\ (K)\end{tabular}} & \textbf{\begin{tabular}[c]{@{}c@{}}LUT \\ (K)\end{tabular}} & \textbf{DSP48E} & \textbf{Latency} & \textbf{\begin{tabular}[c]{@{}c@{}}Clock-Cycle \\ Time (ns)\end{tabular}} & \textbf{\begin{tabular}[c]{@{}c@{}}Throughput\\ (Hz)\end{tabular}} & \textbf{MSE} & \textbf{LSD} \\ \hline
\textbf{Design  /  Available Resources} & 94.5 & 360 & 3456 & 1728 & 12288 &  &  &  &  &  \\ \hline
FloatingPointBaseline & 93 (98\%) & 144 (40\%) & 35 (1\%) & 86 (5\%) & 288 (2\%) & 12110989 & 8.83 & 9.4 & 0 & 0 \\ \hline
FloatingPointCQ & 93 (98\%) & 144 (40\%) & 35 (1\%) & 83 (5\%) & 330 (3\%) & 6170104 & 8.83 & 18.4 & 0 & 0 \\ \hline
FloatingPointPipeline & 93 (99\%) & 144 (40\%) & 231 (7\%) & 231 (13\%) & 475 (4\%) & 612952 & 8.88 & 183.7 & 0 & 0 \\ \hline
FixedPointUnrolling & 79 (84\%) & 144 (40\%) & 22 (1\%) & 146 (8\%) & 660 (5\%) & 293914 & 8.75 & 388.8 & 0.006 & 0.104 \\ \hline
FixedPointMME (Best) & 90 (96\%) & 144 (40\%) & 425 (12\%) & 1669 (97\%) & 540 (4\%) & 78275 & 8.66 & 1475.2 & 0.006 & 0.104 \\ \hline
\end{tabular}
}
\caption{Resource Utilization, Performance and Measured Error in generation for each design implementation. The error metrics namely, Mean Squared Error (MSE) and Log-Spectral Distance (LSD) is measured by comparing the generated audio from FPGA implementations against audio generated from corresponding GPU implementation. The percentages reported indicate percentage of resources utilized by the design. }
\label{fig:resultsfinal}
\end{table*}

\squeezeup
\subsection{\textbf{Network Description Module}}
In this module, we implement the overall architecture of our network as a stack of dilated conventional layers and perform queue update operations followed by a fully connected layer. This module instantiates the corresponding function for each network layer and manages the layer inter-connections. Since each layer is independently instantiated, we can use custom dilation, channels and parallelization parameters for each layer. After the last fully connected layer, to make audio generation deterministic we use arg-max sampling. This allows us to bypass the final softmax layer since we can directly apply the arg-max function on the output of our final fully connected layer.

\begin{figure*}[htp]
    \centering
     \includegraphics[width=1.8\columnwidth]{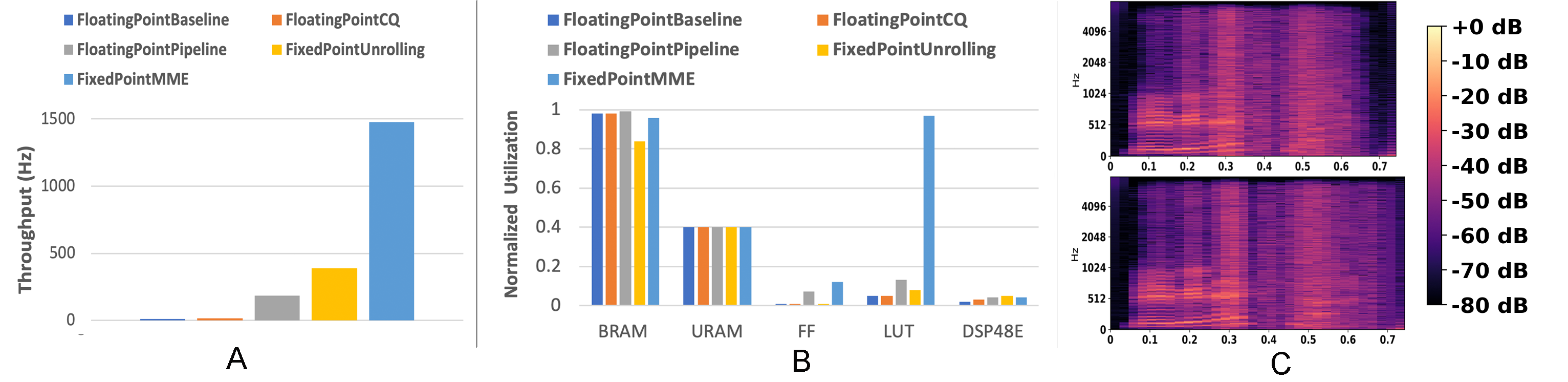}
    \caption{\textbf{A:} Throughput (Number of Samples generated per second) of different designs. \textbf{B:} Normalized Resource Utilization of different designs. \textbf{C:} Log-Spectrograms of the 2-second audio generated from the TensorFlow implementation (top) and FPGA FixedPointMME design implementation (bottom).}\label{fig:resourcegraphs}
\end{figure*} 

\squeezeup
\section{Results and Experiments}
In this section, we evaluate the effect of our optimizations, namely cyclic queues, pipelining, loop unrolling and customized matrix multiplication engine, by conducting extensive design space exploration. Our design experiments are synthesized for the Xilinx XCVU13P board using Xilinx Vivado HLS 2017.4. In particular, we discuss the experimental techniques applied to reduce resource utilization and latency of our baseline implementation. We further provide a comprehensive comparison of our best designs with CPU and GPU implemented baselines in terms of throughput and power efficiency. 

\squeezeupt
\subsection{\textbf{Evaluation Metrics}}
To evaluate the accuracy of our implementation we compare the output generated from our FPGA implementation with the golden output generated by the TensorFlow GPU implementation for the same initial seed. We use the following metrics to compare any two audio signals $x_1$, $x_2$ of the same length:
\begin{itemize}
    \item \textbf{Mean Squared Error (MSE)}: The mean squared error (MSE) between any two given signals $x_1, x_2$ is the mean squared error between their representations in time domain as a sequence of floating point numbers. That is,
    $MSE = mean( (x_1 - x_2)^2 )$.
    The MSE losses reported are from the comparison of the entire waveform i.e. the total mean squared error from all 32000 samples.
    \item \textbf{Log-Spectral Distance (LSD)}: The log-spectral distance~\cite{Rabiner} is a commonly utilized metric, obtained as the root mean square error between the normalized log-spectra of given signals. Given two signals $x_1$, $x_2$, we calculate log-spectral distance between them as follows:
    \begin{equation}
    \begin{multlined}
        ps_1 = (abs(stft(x_1)))^2 \\
        ps_2 = (abs(stft(x_2)))^2 \\
        ls_1 = normalize( log(ps_1) )\\
        ls_2 = normalize( log(ps_2) )\\
        LSD = RMSE( ls_1, ls_2)
    \end{multlined}
    \end{equation}
    Here $ps_1$, $ps_2$ are the power spectra and $ls_1$, $ls_2$ are the normalized log spectra of signals $x_1$, $x_2$ respectively. The normalization is performed across all frequencies in the log spectrograms.
    
    \item \textbf{Qualitative Evaluation}: Along with the quantitative results, we also provide log-spectrogram visualizations of the audio signal generated using our FPGA implementation and the golden-output audio signal generated from the TensorFlow implementation in Figure \ref{fig:resourcegraphs} (c).
\end{itemize}


\squeezeup
\subsection{\textbf{Design Space Exploration}}
We implement the following designs to study the effect of various optimization techniques in isolation and in combination with other techniques. The resource utilization, performance (throughput) and error in the generated audio, for each of the following designs have been reported in Table \ref{fig:resultsfinal}. Throughput measures the number of audio samples generated per second by our implementation of an autoregressive model. Note that one second of audio contains 16000 samples if audio is sampled at 16KHz. 

\subsubsection{Baseline Floating Point Implementation \textit{(FloatingPointBaseline)}}
The baseline design of our network is comprised of modules to implement the basic functionality of each layer, queue, initialization of weights from stored data files and forward propagation. We use a array-shifting implementation of queue which results in a fairly high latency as shown in Table \ref{fig:resultsfinal} because of the very long queues (length = $8192$ and $4096$) in the $\mathbf{13^{th}}$ and $\mathbf{14^{th}}$ layers of our design. For matrix vector multiplication we use simple for loops without any optimization.

\subsubsection{Floating Point + Cyclic Queue \textit{ (FloatingPointCQ)} }
In this design, we replace our shifting based queue implementation with a cyclic queue implementation that uses dynamic indexing to produce the same effect as push and pop operations. This helps reduce latency substantially since shifting operations in the longer queues was the bottleneck in our baseline design. The resource utilization however, stays almost the same as our baseline design. 

\subsubsection{Floating Point + Cyclic Queue + Pipelining \textit{  (FloatingPointPipeline)} }
In this design, we modify the above design and add pipelining pragma in the dot product computation and queue update operations. Pipelining the above design helped increasing the throughput substantially at the cost of higher resource utilization.

\subsubsection{Fixed Point + Cyclic Queue + Unrolling \textit{(FixedPointUnrolling)} }
Including both Cyclic Queue and Pipelining optimization, we switch to fixed point operations from floating point operations. Since the order of magnitude of our kernels, inputs, activations and outputs is nearly the same, we keep a common data-type across all of them. 
After some experimentation, we found that Loop Unrolling outperforms pipelining in terms of both resource utilization and throughput for fixed point data-types. We use \textit{loop unrolling factor = 8} for the inner loop of our dot product and also the queue update operations. We observe a trade-off between precision and resource utilization for different fixed point bit width and chose \texttt{ap\_fixed<27,8>} (8 bits for the integer and 19 bits for the fractional part)since it gives reasonable MSE under the constraints of resources.

\subsubsection{Fixed Point + Matrix Multiplication Engine \textit{(FixedPointMME - Best)} }
For our best design, we use fixed-point implementation in a parallelized approach to convert layer computations into multiple MAC operations (refer to Section \ref{sec:matmul} for details). For the first dilated convolution layer we set \textit{num\_parallel\_out} and \textit{num\_parallel\_in} as 1 since the number of input channels is just 1. For all other layers, including the fully connected layer we set \textit{num\_parallel\_out} as 8 and \textit{num\_parallel\_in} as 4 to get the best throughput under the constraint of available resources.

\begin{table}[htp]
\centering
\resizebox{0.99\columnwidth}{!}{%
\begin{tabular}{|l|c|c|}
\hline
\multicolumn{1}{|c|}{\textbf{Implementation}} & \textbf{\begin{tabular}[c]{@{}c@{}}Time (in seconds) \\ for 1-Second \\ Audio Generation\end{tabular}} & \textbf{\begin{tabular}[c]{@{}c@{}}Power\\ (W)\end{tabular}} \\ \hline
\rowcolor[HTML]{EFEFEF} 
CPU (Numpy) & 732 &  \\
\rowcolor[HTML]{EFEFEF} 
GPU - NVIDIA Titan Xp  (TensorFlow) & 120 & 70 \\
\rowcolor[HTML]{EFEFEF} 
GPU - NVIDIA Tesla V100  (TensorFlow) & 85 & 66 \\
FPGA- FloatingPointPipeline & 87 & 10.2 \\
FPGA- FixedPointUnrolling & 41 & 7.6 \\
\textbf{FPGA- FixedPointMME (Best)} & 11 & 23 \\ \hline
\end{tabular}%
}
\caption{Power Consumption and Wall-Clock time required when generating 1-second audio for different implementations. 
}
\label{fig:cpugpu}
\end{table}

\squeezeup
\subsection{\textbf{Performance and Power Analysis}}\label{sec:power}
Table~\ref{fig:cpugpu} illustrates the performance and power consumption for our implemented designs and a highly optimized CPU and GPU implementation.
We benchmark the optimized Tensorflow implementation of Fast-Wavenet on two GPUs: NVIDIA TITAN Xp and Nvidia Tesla V100. The CPU implementation is the NumPy inference program written by us and optimized fully. We measure the power consumption for the GPU benchmarks using the NVIDIA power measurement tool (\textit{nvidia-smi}) running on \textit{Linux} operating system which is invoked during program execution. For our FPGA implementations, we synthesize our designs using Xilinx Vivado v2017.4. We then integrate the synthesized modules accompanied by the corresponding peripherals into a system-level schematic using Vivado IP Integrator. The frequency is set to 150 MHz and power consumption is estimated using the synthesis tool. 

As shown, our best FPGA implementation achieves $\mathbf{11\times}$ speed-up in audio generation while being $\mathbf{3\times}$ more power efficient as compared to NVIDIA Titan Xp GPU. As compared to a NumPy based CPU implementation, our best design is $\mathbf{66\times}$ faster.

\section{Prior Works on Accelerating DNNs for FPGAs}
Prior works have made significant efforts in compressing Deep Neural Networks (DNNs) to support fast energy-efficient applications. However, recent research on DNNs is still increasing the depth of models and introducing new architectures, resulting in higher number of parameters per network and higher computational complexity. Other than CPUs and GPUs, FPGAs are becoming a platform candidate to achieve energy efficient neural network computation~\cite{zhang2015optimizing,ovtcharov2015accelerating,suda2016throughput,sharma2016high,samragh2017customizing,shea2018scalenet,rnnaccelerator}. Equipped with the necessary hardware for basic DNN operations, FPGAs are able to achieve high parallelism and utilize the properties of neural network computation to remove unnecessary logic. Algorithm explorations also show that neural networks can be simplified to become more hardware friendly without sacrificing the accuracy of the model. Therefore, it has become possible to achieve increased speedup and higher energy efficiency on FPGAs compared to CPU and GPU platforms~\cite{surveyoffpga,zhang2015efficient} while maintaining state-of-the-art accuracy. 
Prior efforts have been made in FPGA acceleration of speech recognition, classification and language modelling using Recurrent Neural Networks\cite{rnnaccelerator,fpga_lowpower,rnnlanguage}; however the challenges in \textit{generation} of sequences with \textit{long-term dependencies}, particularly in audio domain have not been addressed. 

\squeezeup
\section{Conclusion} \label{sec:discussion}
We present the first accelerator platform for deep autoregressive convolutional neural networks. While prior works have proposed algorithms for making the inference of such networks faster on GPUs and CPUs, they do not exploit the potential parallelism offered by FPGAs. We develop a systematic approach to accelerate the inference of WaveNet based neural networks by optimizing their fundamental computational blocks and utilizing only on-chip memory. We demonstrate the effectiveness of using FPGAs for fast audio generation by achieving a significant speed-up over prior efforts on CPU and GPU based implementations.

\bibliographystyle{IEEEtran}
\bibliography{myref.bib}



\end{document}